\documentclass[mathleft
]{an}
\usepackage{graphicx}
\usepackage{times}
\usepackage{natbib}
\bibpunct{(}{)}{;}{a}{}{,}
\def\note #1]{{\bf #1]}}

\begin{document}

\Pagespan{789}{}
\Yearpublication{2006}%
\Yearsubmission{2005}%
\Month{11}%
\Volume{999}%
\Issue{88}%

\title{Seismological challenges for stellar structure}

\author{J{\o}rgen Christensen-Dalsgaard\inst{1}\fnmsep
\thanks{\email{jcd@phys.au.dk}\newline}
}
\titlerunning{Seismological challenges for stellar structure}
\authorrunning{J{\o}rgen Christensen-Dalsgaard}
\institute{
Department of Physics and Astronomy, Building 1520, Aarhus University,
8000 Aarhus C, Denmark}

\received{????}
\accepted{????}
\publonline{later}


\abstract{%
Helioseismology has provided very detailed information about the solar interior,
and extensive data on a large number of stars, although at less detail,
are promised by the ongoing and
upcoming asteroseismic projects.
In the solar case there remain serious challenges in
understanding the inferred solar structure,
particularly in the light of the revised
determinations of the solar surface composition.
Also, a secure understanding of the origins
of solar rotation as inferred from helioseismology,
both in the radiative interior and in the
convection zone, is still missing.
In the stellar case challenges are certain to appear as the
data allow more detailed inferences of the properties of stellar cores.
Large remaining uncertainties in modelling concerns the properties
of convective cores and other processes that may cause mixing.
As a result of developing asteroseismic signatures addressing these and other
issues, we can look forward to a highly challenging, and hence exciting,
era of stellar astrophysics.
}

\maketitle

\section{Introduction}

%
The present is a period of dramatic evolution of the study of stellar structure
and evolution, in large part owing to the possibilities offered by the
rapidly expanding data on stellar oscillations.
The resulting frequencies are expected to provide stringent constraints on
stellar properties, including the structure of stellar interiors, 
and hence challenge our understanding of stellar internal physics and
stellar evolution.
An obvious requirement for this is that observations are made of a sufficiently
broad sample of stars, with sufficient sensitivity and of sufficient duration
to secure the required frequency precision.
Such observations are now resulting from the CoRoT and Kepler space missions
\citep[e.g.,][]{Baglin2006, Boruck2010},
and further improvements are promised by dedicated ground-based facilities
for radial-velocity observations \citep[e.g.,][]{Grunda2009}.
However, this is not all: we must also be able to analyse the basic observations
to extract reliable frequencies, as well as to identify the observed modes.
Also, stellar modelling must be sufficiently reliable and accurate that 
the assumptions about stellar internal physics are faithfully reflected in
the frequencies computed from the resulting models.
Although substantial progress is being made both in the data analysis and
the modelling, it is probably fair to say that further efforts are required 
in these directions.

In this brief review I discuss some of the issues of observations and 
modelling, including those where seismic challenges are expected.
The emphasis will largely, but not exclusively, be on stars showing
solar-like oscillations.
Further discussion of some of these issues was presented, for example,
by \citet{Christ2010},
while an extensive overview of asteroseismology was provided by
\citet{Aerts2010}.

\section{The observational situation}

%
The past decade has seen extensive efforts on ground-based observations
of solar-like oscillations \citep[for a review, see][]{Kjelds2004},
culminating in a large coordinated set of observations of Procyon 
\citep{Arento2008, Beddin2010}.
Large campaigns, with important asteroseismic potential,
have also been carried out for other types of pulsating stars
\citep[e.g.,][]{Breger1995, Kawale1995, Kurtz2002, Aerts2004, Handle2004}.

A major breakthrough has been the development of space-based asteroseismology,
starting with the serendipitous use of the WIRE satellite
\citep[e.g.,][]{Buzasi2005}.
The Canadian micro-satellite MOST \citep[e.g.,][]{Walker2003}
probably cannot quite reach
the low level of noise required for the observation of solar-like oscillations
in main-sequence stars \citep{Matthe2004, Beddin2005},
but has yielded high-quality data for a substantial number of other types 
of pulsating stars \citep[e.g.,][]{Saio2006, Matthe2007}.
The French-led CoRoT satellite, on the other hand, has resulted 
in a break-through in the study of solar-like oscillations
\citep[e.g.,][]{Michel2008, DeRidd2009},
as well as for more classical pulsating stars
\citep[e.g.,][]{Degroo2009,Porett2009}.
A major strength is the availability of nearly continuous
observations extending over five months.
Even more extensive observations are promised by the NASA Kepler mission,
launched in March 2009 \citep{Koch2010, Kjelds2010}.
Even the early analysis of the first month of Kepler data has demonstrated
the outstanding quality of these data \citep{Gillil2010},
and it is expected that, following a year-long survey phase,
a substantial number of stars will be observed for the full, or a large
fraction of, the remaining duration of at least 2 1/2 years of the mission.

Very extensive asteroseismic data of high quality, as an integral part
of the study of extra-solar planetary systems, will be obtained by the
ESA PLATO mission, to be launched around 2018 if it is finally selected 
for implementation \citep{Catala2009, Zima2010}.

Since the intrinsic stellar `noise' from granulation, activity, etc., is 
substantially higher relative to the oscillations in intensity than in
velocity \citep{Harvey1988},
the ultimate precision of asteroseismic investigations requires 
radial-velocity observations.
These can be carried out from the ground, but the necessary sensitivity,
continuity and duration require dedicated facilities, either in a network of
telescopes at low or intermediate latitudes or from Antarctica.
A network of telescopes is the goal of the SONG project
\citep{Grunda2009, Grunda2010a, Grunda2010b},
while the SIAMOIS project \citep{Mosser2008}
seeks to establish a facility for radial-velocity
asteroseismology at the Concordia station on Dome C in Antarctica.


Asteroseismic use of the observed frequencies requires that the observed
oscillations be identified with modes of stellar models.
In the case of solar-like oscillations, the determination of
the spherical-harmonic degree of the modes can in many cases be
accomplished through identification of the acoustic-mode asymptotic
structure (see Eq.\,(\ref{eq:pasymp}) below);
however, the example of Procyon shows that this this does not always
provide definite results \citep{Beddin2010, Dogan2010}.
In stars showing low-order oscillations excited through a heat-engine
mechanism no similarly simple frequency structure is typically observed;
here it is crucial to determine the degree (and possibly the azimuthal order)
of the modes observationally, e.g., by supplementing space observations
by ground-based campaigns \citep{Uytter2010}.

\section{Asteroseismic diagnostics}

%
The oscillation spectra of relatively unevolved stars, including the Sun,
are essentially separated into high-frequency acoustic (or p) modes and
low-frequency buoyancy-driven internal gravity (or g) modes.
For solar-like oscillations the observations are of high-order p modes,
approximately satisfying an asymptotic relation of the form
\begin{equation}
\nu_{nl} = \Delta \nu ( n + l/2 + \epsilon_0) - l(l+1) D_0 \; ,
\label{eq:pasymp}
\end{equation}
where $n$ is the radial order, and $l$ the spherical-harmonic degree,
of the mode.
This is characterized by a large frequency separation 
$\Delta\nu = (2 \int_0^R {\rm d}r/c)^{-1}$ determined
by the integral of the inverse sound speed $c$ over the stellar radius;
$\Delta \nu$ essentially scales as $\bar \rho^{1/2}$,
where $\bar\rho \propto M/R^3$ is the mean density of the star,
$M$ and $R$ being its mass and radius.
The term in $D_0$ gives rise to a small frequency separation
$\delta \nu_{nl} = \nu_{nl} - \nu_{n-1 \, l+2}$
that is sensitive to the variation of the sound speed in the core
of the star, which, for main-sequence stars,
in turn changes with the core composition and hence the age of the star.
High-order g modes are also observed in several cases, such as 
$\gamma$ Dor and slowly pulsating B stars, as well as in white dwarfs.
For these the periods $\Pi_{nl}$ are approximately uniformly spaced, satisfying
\begin{equation}
\Pi_{nl} = {\Pi_0 \over \sqrt{l(l+1)}} (n + \epsilon_{\rm g}) \; ,
\label{eq:gasymp}
\end{equation}
where the basic period spacing $\Pi_0$ depends on the buoyancy frequency
in the stellar interior.

For evolved stars the buoyancy frequency, and hence the g-mode frequencies,
reach very high values in the stellar core, 
leading to an overlap between the p- and g-mode spectra.
This causes avoided crossings between the frequencies of the two classes
as the star evolves, of potentially very high diagnostic value
\citep{Deheuv2010}.
In red giants the extremely dense spectrum of g modes very substantially
complicates the frequency computation 
and to some extent the interpretation of observed frequencies
\citep[e.g.,][]{Dziemb2001, Christ2004, Dupret2009}.

For spherically symmetric (and hence non-rotating) stars the frequencies are
independent of the azimuthal order $m$ of the modes. 
Rotation lifts this degeneracy; 
for slow rotation the resulting frequencies are of the form
\begin{equation}
\nu_{nlm} \simeq \nu_{nl0} + 2 \pi m \beta_{nl} \langle \Omega \rangle \; ,
\label{eq:rotsplit}
\end{equation}
where $\beta_{nl}$ is a constant that in most cases is close to one,
and $\langle \Omega \rangle$ is an average, depending on the mode,
of the stellar angular velocity $\Omega$.


The initial analysis of observed frequencies of solar-like oscillations
is typically based on the parameters characterizing the asymptotic
behaviour (Eq.\,\ref{eq:pasymp}), in terms of the large and small 
frequency separations.
These provide a measure of the mass and evolutionary state of the star
\citep[e.g.,][]{Christ1984, Christ1988, Ulrich1986}, although
subject to other uncertainties in the parameters characterizing
the star \citep{Gough1987}.
A full use of the observations evidently requires analysis of the individual
frequencies,
typically through a $\chi^2$ fit involving the frequencies and possibly
other observed properties of the star, such as effective temperature,
luminosity, etc.\ \citep[e.g.][]{Miglio2005}.
An important issue is to ensure that the inferred solution provides 
the optimum fit to the observations, amongst the parameters characterizing
the model, rather than a local minumum in $\chi^2$.
This can either be achieved through the use of extensive
multidimensional precomputed model grids \citep[e.g.,][]{Guenth2004}
or through efficient fitting procedures involving recomputation of models,
such as the genetic algorithm \citep{Metcal2009}.
Once an approximation to the best solution has been found,
it can be further refined through a local fit around this approximation;
this can efficiently be carried out through singular-value decomposition
of the relation between corrections to the model parameters and the misfit
to the observations, which furthermore provides useful information about
the properties of the solution \citep{Brown1994, Creeve2007}.


Such fits evidently lead to the determination of the parameters
characterizing a given set of model calculations but do not in themselves
provide a challenge to stellar structure.
A challenge may show up, however,
as an inconsistency between the resulting models
and other, independent, observations of the star;
this underlines the importance of including other types of stellar observations
in the analysis of asteroseismic data \citep{Molend2010}.
A formal expression of the level of challenge to the model calculations 
is obtained if, as is often
the case, the model frequencies and other properties differ significantly
from the corresponding observed values
(assuming, of course, that a reliable measure of the uncertainties
in the latter is available).
In this case more sophisticated types of analysis are required to
determine the origin of the differences, very likely leading to true
challenges to the theory of stellar evolution.
Such analyses may concentrate on specific features in the star,
such as the borders of convective regions (see Section 6 below).
Alternatively, it may be possible in a limited way to carry out inverse
analyses to infer the differences between the model and stellar structure
\citep[e.g.,][]{Basu2002, Roxbur2004}.



\section{Microphysical challenges}

%
An important aspect of asteroseismology is the potential to use stars
as `laboratories' for the study of the properties of matter under extreme
conditions.
In the solar case, striking tests have been made of the treatment of
the thermodynamical state of matter, certainly challenging existing
formulations \citep[e.g.,][]{Elliot1998, Basu1999}.
Interesting evidence has been found for crystallization in pulsating
white dwarfs \citep[e.g.,][]{Metcal2004} which may help elucidating these
processes, of great importance to the cooling timescale of white dwarfs.
On the other hand, thermodynamic effects on solar-like oscillations are
probably too subtle to be detectable with present observations.

Historically, asteroseismology has played an important role in the 
development of modern computations of stellar opacity, following
Simon's plea \citep{Simon1982}.
The current discrepancies between the helioseismically inferred solar 
structure and the structure of solar models
\citep[for a review, see][]{Basu2008},
following the revised determinations of solar abundances \citep{Asplun2009},
certainly represent a serious challenge to stellar modelling;
as noted by, for example, \citet{Bahcal2005} and \citet{Christ2009} this
may be an indication that further opacity revisions are required.
Interestingly, discrepancies between the frequency range of modes
observed in hot stars, and the range of
unstable modes in models of these stars,
may also indicate a need for opacity revisions
\citep[e.g.,][]{Jeffer2007, Miglio2007, Dziemb2008}.
A closely related issue concerns the apparent need for radiative
levitation of iron-group elements to explain the excitation of oscillations
in subdwarf B stars \citep{Fontai2003, Fontai2006}.

%

\section{Near-surface problems}

%
A major challenge to the modelling of cool stars and their oscillations
is the treatment of near-surface convection.
In addition to the resulting temperature structure, usually obtained by
means of a simplified parameterized treatment such as the mixing-length
formulation, the structure of the star is undoubtedly affected by
`turbulent pressure', i.e., the momentum transport resulting from convection;
this is generally ignored in computations of stellar models.
Also, the perturbations to the convective flux and turbulent pressure
are at best included in a simplified, and rather uncertain, manner 
in the computation of the oscillation properties
\citep[see][for further details]{Houdek2010}.


It is striking that most of the difference between observed and computed
frequencies of the Sun does indeed arise in the superficial layers of the
model \citep{Christ1996}, very likely caused by such errors in the treatment
of convection and possibly in failures to model properly the solar atmosphere.
These effects increase rapidly with increasing frequency and hence are
less important in stars such as $\delta$ Scuti stars where generally modes
of relatively low order are observed.
However, they obviously affect the interpretation of observed solar-like
oscillations.
It should be noted that owing to their strong frequency dependence also
the large frequency separation is substantially changed by the near-surface
effects, and even the small frequency separation is significantly
affected.
It was noted, however, by \citet{Roxbur2003} that appropriate ratios between
the small and large separations are virtually insensitive to the outer
layers of the star and hence provide good diagnostics of the properties
of stellar cores \citep[see also][]{Oti2005}.


It is obviously highly desirable to use the full information in the observed
frequencies and hence to carry out analysis of the individual frequencies,
rather than the separation ratios.
This requires understanding of, or correction for, the near-surface effect.
For stars that are not too different from the Sun it may be a reasonable
{\it ansatz} to assume that the functional form of the effects, e.g.
measuring the frequencies in units of the acoustical cut-off frequency,
is similar to the effect in the Sun which can be determined through 
analysis of the frequencies of modes over a broad range of degree, as 
available in the solar case.
\citet{Kjelds2008} assumed a power-law behaviour of the frequency correction,
obtaining the exponent from the solar data, and demonstrated that the
resulting correction yielded reasonable results for a few solar-like stars.
Application to stars observed by Kepler, and to Procyon, has met with
mixed success \citep[e.g.,][]{Dogan2010}.
However, it is likely that observations with Kepler of a broad range of
solar-like stars, in what has been termed `ensemble asteroseismology',
will yield important insight into these effects and their dependence on
stellar properties.

An important goal is clearly to achieve more realistic modelling of 
these effects.
Detailed and relatively realistic radiation hydrodynamical models 
of the outer parts of the solar convection zone are in fact possible
\citep[for a review, see][]{Nordlu2009};
unlike most of the star the dynamical and thermal timescales are comparable
in this region and hence these processes can be treated fully.
Using average properties of the resulting models to represent the outer
layers of full models does yield a substantial improvement in the
agreement between computed and observed solar oscillation frequencies
\citep[e.g.,][]{Rosent1999, Li2002}.
This suggests that the effect of turbulent pressure in the equilibrium 
model may play a significant role.
In principle it may become possible also to study the convective effects
on stellar pulsations through such hydrodynamical simulations, although
longer simulation runs than currently available will likely be
required to isolate the oscillations of the simulation box and 
determine the frequencies and other properties with sufficient accuracy.
One may hope from such simulations to obtain insight that can then inform
simpler treatments that may realistically be included in extensive
computations of stellar oscillations.

\section{Borders of convective regions}

%
Except in extreme stages of stellar evolution
the bulk of stellar convection zones is very nearly adiabatically stratified
and fully mixed.
This represents no significant problems for stellar modelling.
However, the borders of convective regions are seats of great uncertainty.
It is evident that convective motion does not stop at the edges of
convectively unstable regions, yet the extent and nature of the motion
in the adjacent stable region are highly uncertain
\citep[e.g.,][]{Maeder1975, Zahn1991}.
Also, in regions of varying chemical composition additional uncertainty
may arise in {\it semi-convective regions} that are formally stable to
convection but unstable to growing oscillatory motion
\citep[see][for an overview]{Kippen1990}.
Related problems occur in models with growing convective cores
\citep{Popiel2005, Montal2007}.
The uncertainty associated with these processes has an important effect on
possible mixing, particularly outside convective cores, affecting the resulting
composition profile and hence the subsequent evolution of the stars.

The rapid variations in the density gradient associated with the transition
to convective stability and possibly composition variations at the
edges of convective regions give rise to glitches in the acoustic and
gravity-wave properties which affect stellar oscillations.
Specifically, they introduce an oscillatory behaviour in the frequencies,
as a function of frequency, arising from the varying phase of the
oscillation at the location of the glitch, 
and causing departures from the simple asymptotic behaviour
in Eqs\,(\ref{eq:pasymp}) or (\ref{eq:gasymp}).
For acoustic modes \citet{Gough1990} proposed considering the
second difference 
$\Delta_2 \nu_{nl} = \nu_{n-1 \, l} - 2 \nu_{nl} + \nu_{n+1 \, l}$
which obviously vanishes if Eq.\,(\ref{eq:pasymp}) is valid.
Effects of acoustic glitches are clearly seen in observed solar
oscillation frequencies and have been used to constrain conditions at the
base of the solar convective envelope
\citep[e.g.,][]{Basu1994, Montei1994, Roxbur1994}.
A similar analysis is in principle possible based on just the low-degree
modes observed in solar-like stars \citep[e.g.,][]{Montei2000}.
It should also be noted that other features in the star may cause
acoustic glitches;
particularly important is the effect on the sound speed of the ionization
of helium which gives rise to a signature that has been used to determine
the solar helium abundance.
A careful analysis of the effect of such acoustic glitches was reported
by \citet{Houdek2007}.

In evolved main-sequence stars, particularly with convective cores,
the variation in the core of the hydrogen abundance gives rise to a rapid
variation of the sound speed with potentially observable effects on the
oscillation frequencies.
This is particularly dramatic in stars with growing convective cores,
as typically found for masses below around $1.5 \, {\rm M}_\odot$ where 
a discontinuity in composition, density and hence sound speed results
if diffusion is neglected.
This behaviour, including asteroseismic diagnostics, was analysed by
\citet{Popiel2005} and \citet{Cunha2007};
further analyses of the resulting diagnostics are presented by
\citet{Cunha2010} and \citet{Branda2010}.

For g modes the relevant property are sharp variations in the buoyancy
frequency, giving rise to departures from the uniform period spacing
in Eq.\,(\ref{eq:gasymp}).
This has been used extensively to characterize the properties of 
pulsating white dwarfs \citep[for a review, see][]{Fontai2008}.
The variation in the buoyancy frequency outside convective cores
also has a substantial effect on g modes in main-sequence stars, particularly
visible when, as in the case of slowly pulsating B~stars, high-order 
modes are observed.
\citet{Miglio2008} carried out a careful analysis of the resulting 
signatures in the period spacings.
Remarkably, this behaviour was recently observed in CoRoT data on a
B~star \citep{Degroo2010a, Degroo2010b},
promising very valuable constraints on
the mixing outside convective cores in such stars.

%
%
%

\section{Effects of rotation}

%
In the solar case the availability of a huge number of modes, spanning
a large range in degree and azimuthal order,
has allowed detailed inferences of the internal rotation of the Sun
\citep[see][for reviews]{Thomps2003, Howe2009}. 
The results represented serious challenges, so far not entirely
resolved, to the existing models of the evolution of rotation in
stellar interiors.
In particular, the transport of angular momentum from
an initially rapidly rotating solar interior, as is generally assumed,
remains contentious.
Inferences of internal rotation of other solar-like stars would
be extremely helpful in resolving this issue.
Unfortunately, the rotational splittings 
(cf.\ Eq.\,\ref{eq:rotsplit}) for the low-degree acoustic modes observed 
in most solar-like oscillations are predominantly sensitive to rotation in
the outer parts of the star;
however, if compared with a measure of the surface rotation, e.g.,
from variations caused by the rotation of surface features,
$\langle \Omega \rangle$ as inferred from the splitting provides 
an indication of the variation of rotation in the stellar interior.
More detailed information about rotation in the deep interior
will be available from observations of mixed modes, with partial g-mode
character, in evolved stars \citep{Lochar2004}.
Observation of low-order modes are also generally more sensitive
to the internal rotation.
In particular, \citet{Dupret2004} found that the core rotation of
the $\beta$ Ceph star HD\,129929 substantially exceeded the rotation
of the outer parts.

Modelling of rotating stars remains somewhat uncertain
\citep[for a detailed overview, see][]{Maeder2009}.
The dynamical effects of the centrifucal acceleration
are relatively straightforward to include, at least if rotation
is not too rapid.
Far more uncertain is the modelling of the rotationally induced circulation
and other mixing processes, and the related evolution of the internal
rotation profile.
The formulation of \citet{Zahn1992}, with further refinements, has
seen extensive use in stellar modelling, with some successes, e.g.,
in reproducing observed surface abundances as the result of mixing processes.
However, these models fail to reproduce the helioseismically inferred
solar internal rotation.
This clearly underlines the importance of further observational information
about the evolution of rotation in other stars,
as additional constraints for the modelling.

The rotational splitting in Eq.\,(\ref{eq:rotsplit}) provides a 
clean diagnostic of the internal stellar rotation, but only for slow
rotation.
With faster rotation higher-order terms in $\Omega$ must be included.
Also, the rotational splitting becomes comparable with the frequency
separation between different multiplets $(n,l)$, greatly complicating the
interpretation of the observed spectra.
As reviewed by \citet{Reese2010} the perturbative description, in which
Eq.\,(\ref{eq:rotsplit}) is the lowest order, breaks down for
sufficiently rapid rotation, leading to more complex types of oscillation;
in practice this happens at rotation rates such as to be relevant for many
asteroseismically interesting stars.
Much work is required on the analysis and asteroseismic interpretation 
of observed oscillations in such cases.

%
%
%

\section{How do we proceed?}

%
%
%
It is traditional for a review on asteroseismology to end by asking for
more and better data. 
It is certainly the case that our wishes are being granted; 
the present situation, with CoRoT and Kepler, has been described as
`drinking from a firehose', and the situation will get even more extreme if
PLATO is selected for implementation.
However, in particular the availability of very long timeseries at the highest
possible quality, and for a broad range of stellar types, will be crucial
to investigate the physical processes in stellar interiors, moving beyond
the overall properties of stars.
This is also the strong argument for extended observations in radial velocity
from the ground, as will be provided by the SONG network.

Given the huge effort that is going into the observations, major efforts
are also called for to optimize the analysis of the resulting data.
Even in the case of helioseismology it is far from clear that the data
are utilized optimally \citep{Jeffer2004}.
In asteroseismology challenges are the lower signal-to-noise ratio
than for the best solar data, as well as the potentially more complex
oscillation spectra than for the Sun, in many cases;
also, as in the solar case, the stochastic nature of the excitation must
be taken into account.
For the further interpretation of the data it is important to obtain
statistically meaningful estimates of the oscillation properties.
Such estimates can in principle be obtained from global fits to the power
spectra, constrained by Bayesian priors
\citep[e.g.,][]{Appour2008, Benoma2009, Gruber2009}.
To be fully reliable,
such analyses depend on a thorough understanding of the statistical
properties of the oscillations, stellar background and observing procedure.

In conclusion, how serious are the current seismological challenges
to stellar structure?
There are certainly examples of observations that are not yet understood;
however, with the exception of helioseismology
these are perhaps not quite yet at a level where our fundamental
understanding of stellar structure and evolution has been challenged.
To reach this level must be a serious challenge
for the whole asteroseismic community, from observations, through
data analysis and interpretation, to a reliable confrontation of the results
with the stellar models.
We shall surely meet this challenge!

%
%
%
%

\acknowledgements
I am very grateful to the organizers for an excellent conference 
and a fitting conclusion to HELAS in its present form, in a beautiful
location.
This work was supported by the European Helio- and Asteroseismology
Network (HELAS), a major international collaboration funded by the
European Commission's Sixth Framework Programme.

%


\end{document}